\documentclass[aps,prl,twocolumn]{revtex4}
\pdfoutput=1
\usepackage{graphicx}
\usepackage{multirow}
\usepackage{amssymb}

\begin{document}

\title{Evidence for the \textbf{$E_{2u}$} model of the superconducting order parameter\\ in UPt$_3$ from Josephson interferometry}

\author{J. D. Strand}
\email{strand2@illinois.edu}
\author{D. J. Van Harlingen}
\email{dvh@illinois.edu}
\affiliation{University of Illinois at Urbana-Champaign}

\author{J. B. Kycia}
\email{jkycia@scimail.uwaterloo.ca}
\author{W. P. Halperin}
\email{w-halperin@northwestern.edu}
\affiliation{Northwestern University}

\date{\today}

\begin{abstract}
\noindent We present data on the modulation of the critical current with applied magnetic field in UPt$_3$--Cu--Pb Josephson junctions and SQUIDs.  The junctions were fabricated on polished surfaces of UPt$_3$ single crystals.  The shape of the resulting diffraction patterns provides phase sensitive information on the superconducting order parameter.  Our corner junction data show asymmetric patterns with respect to magnetic field, indicating a complex order parameter, and both our junction and SQUID measurements point to a phase shift of $\pi$, supporting the $E_{2u}$ representation of the order parameter.
\end{abstract}

\maketitle

\section{Introduction}
More than two decades after its discovery\cite{stewart84}, the mechanism of superconductivity in UPt$_3$ is still unknown.  Although UPt$_3$ has many unusual properties which suggest that the order parameter is unconventional, the pairing symmetry has not been unambiguously determined.  Perhaps most unusual is that it exhibits a characteristic double peak in specific heat\cite{fisher89}, indicating two distinct superconducting phases, with an initial transition at $T_{c+} \approx550mK$ and a second transition $T_{c-}\approx500mK$.  Subsequent measurements revealed a third phase at high magnetic field\cite{adenwalla90}.  Transport measurements show power law dependencies at low temperatures, revealing the presence of nodes in the superconducting gap\cite{behnia91,suderow97}.  Muon spin resonance showed signs of spontaneous magnetization, and thus time-reversal symmetry breaking (TRSB) in the low temperature phase\cite{luke93}, but this result has not been reproduced in later measurements\cite{reotier95}. NMR studies of the Knight shift support a triplet pairing mechanism or possibly a singlet state with strong spin-orbit scattering\cite{tou96}.

Numerous models for the pairing symmetry have been put forward to explain this complicated behavior, but the two best candidates are the singlet-state $E_{1g}$ and triplet-state $E_{2u}$ representations of the order parameter\cite{park96,sauls94}.  Both models feature a real order parameter in the high temperature A-phase and a complex order parameter below the second transition in the B-phase.  Various efforts have been made to distinguish between these two theories on the basis of experiment\cite{norman96,graf00,wu02,joynt02}, in particular relying on details of transport properties, but the relatively subtle differences in node structure and gap magnitude have proven difficult to resolve.  Perhaps the clearest difference between these models is the periodicity of phase winding in the order parameter, as seen in Table 1.  A rotation of $90^\circ$ about the c-axis causes a phase shift of $\pi$ in the $E_{1g}$ model, but a phase shift of $2\pi$ in the $E_{2u}$ model. In this paper, we propose to distinguish between these two models by detecting this difference in phase with Josephson interferometry.

\begin{table}
\begin{tabular}{|c|c|c|}

\hline
& A-Phase & B-Phase\\
\hline
\multirow{8}{*}{$E_{1g}$} & 
		\multirow{7}{*}{\includegraphics[width=.16\textwidth]{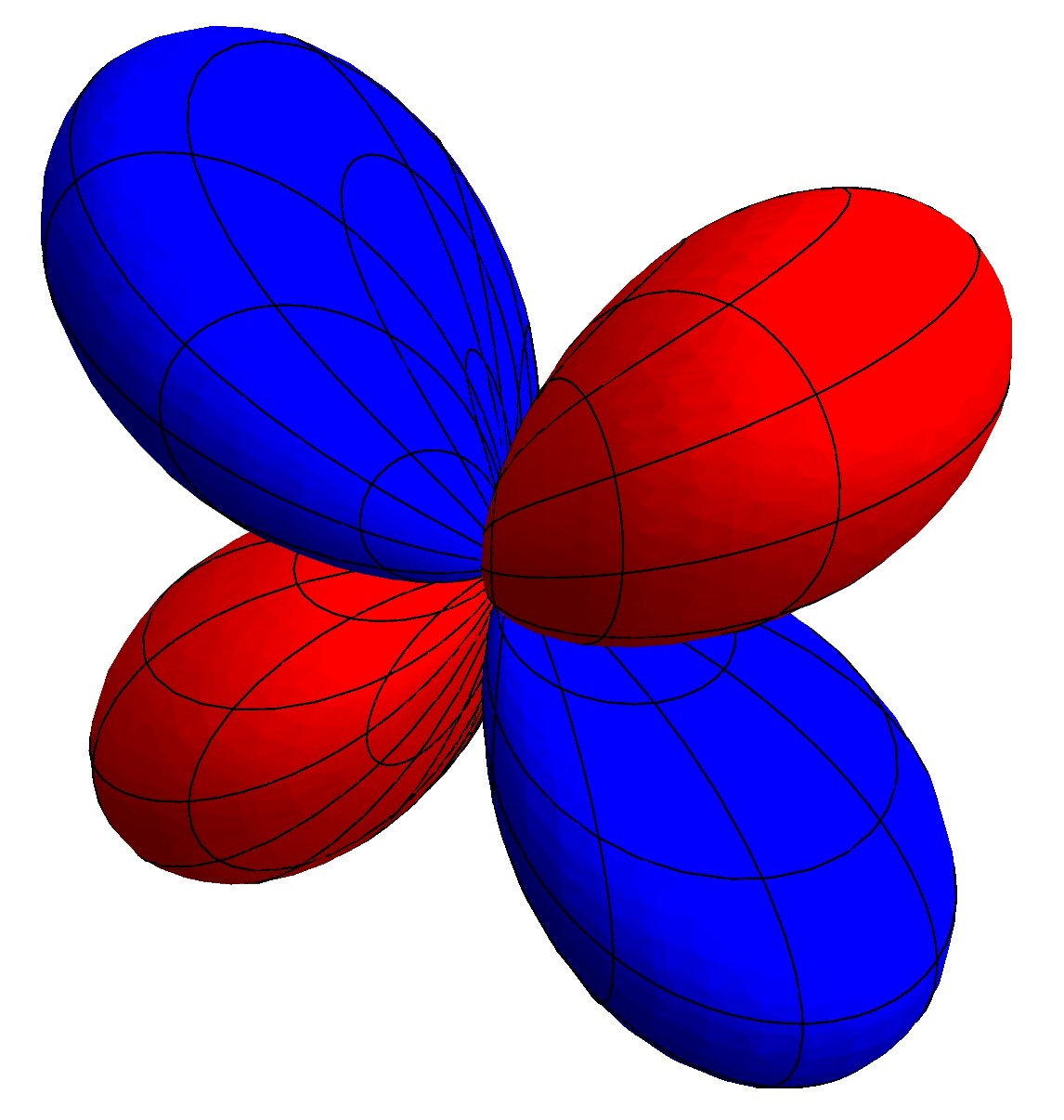}} & 
		\multirow{7}{*}{\includegraphics[width=.15\textwidth]{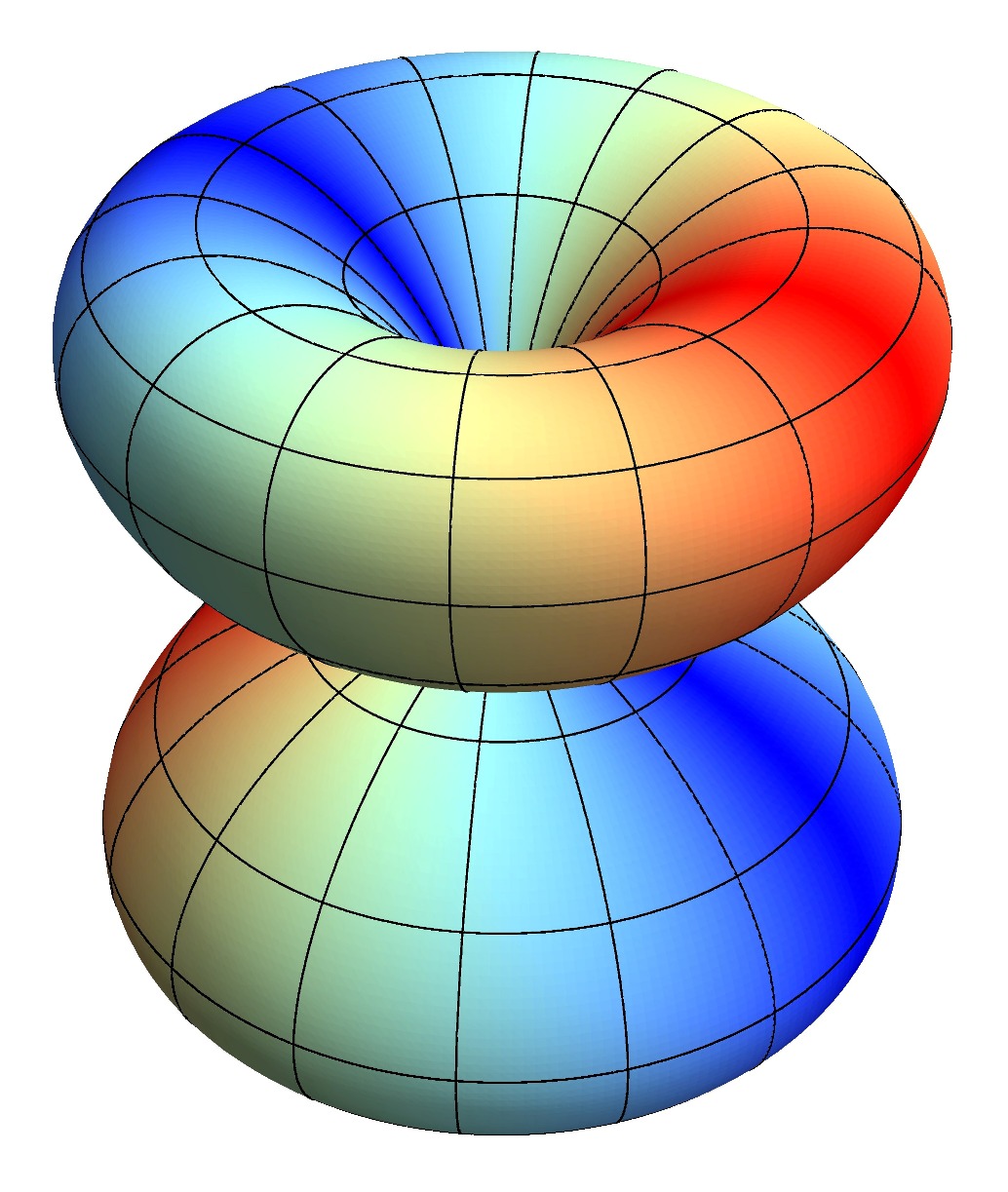}} \\
& & \\
& & \\
& & \\
& & \\
& & \\
& & \\
& $\Delta(k)=\Delta(T)k_xk_z$ & $\Delta(k)=\Delta(T)(k_x+ik_y)k_z$\\
\hline
\multirow{8}{*}{$E_{2u}$} &
		\multirow{7}{*}{\includegraphics[width=.16\textwidth]{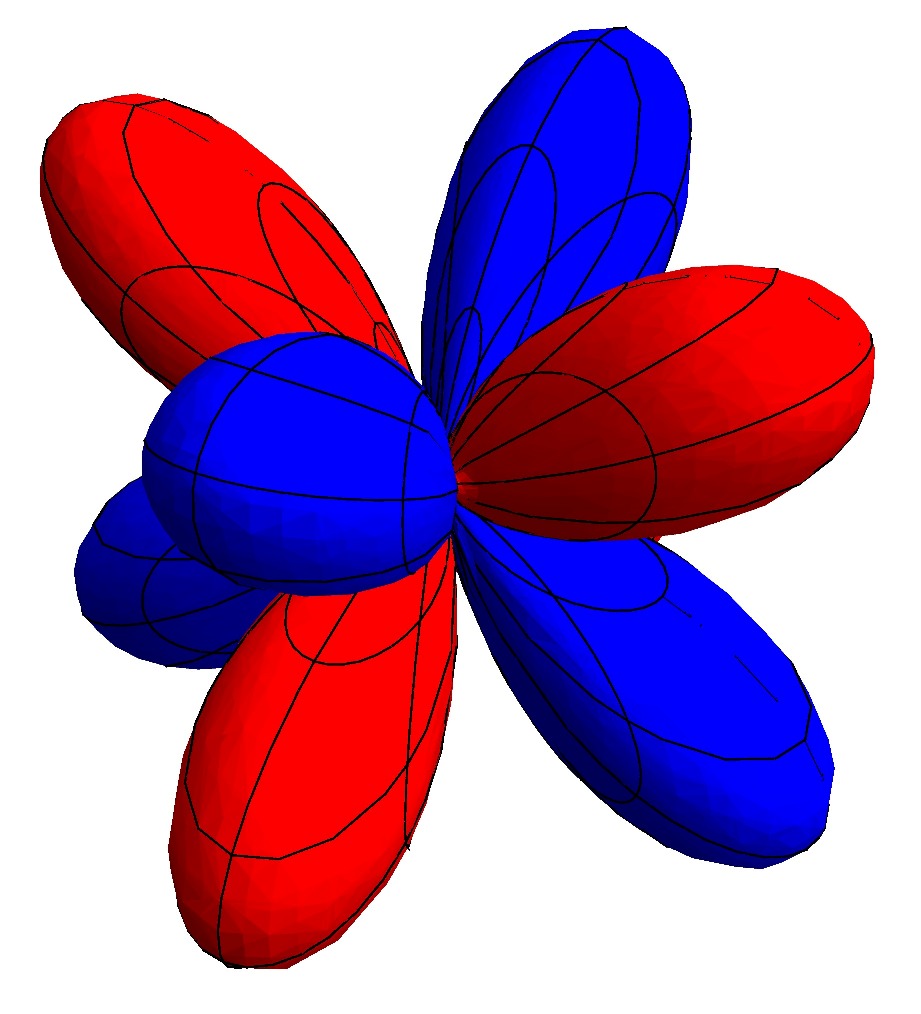}} & 
		\multirow{7}{*}{\includegraphics[width=.16\textwidth]{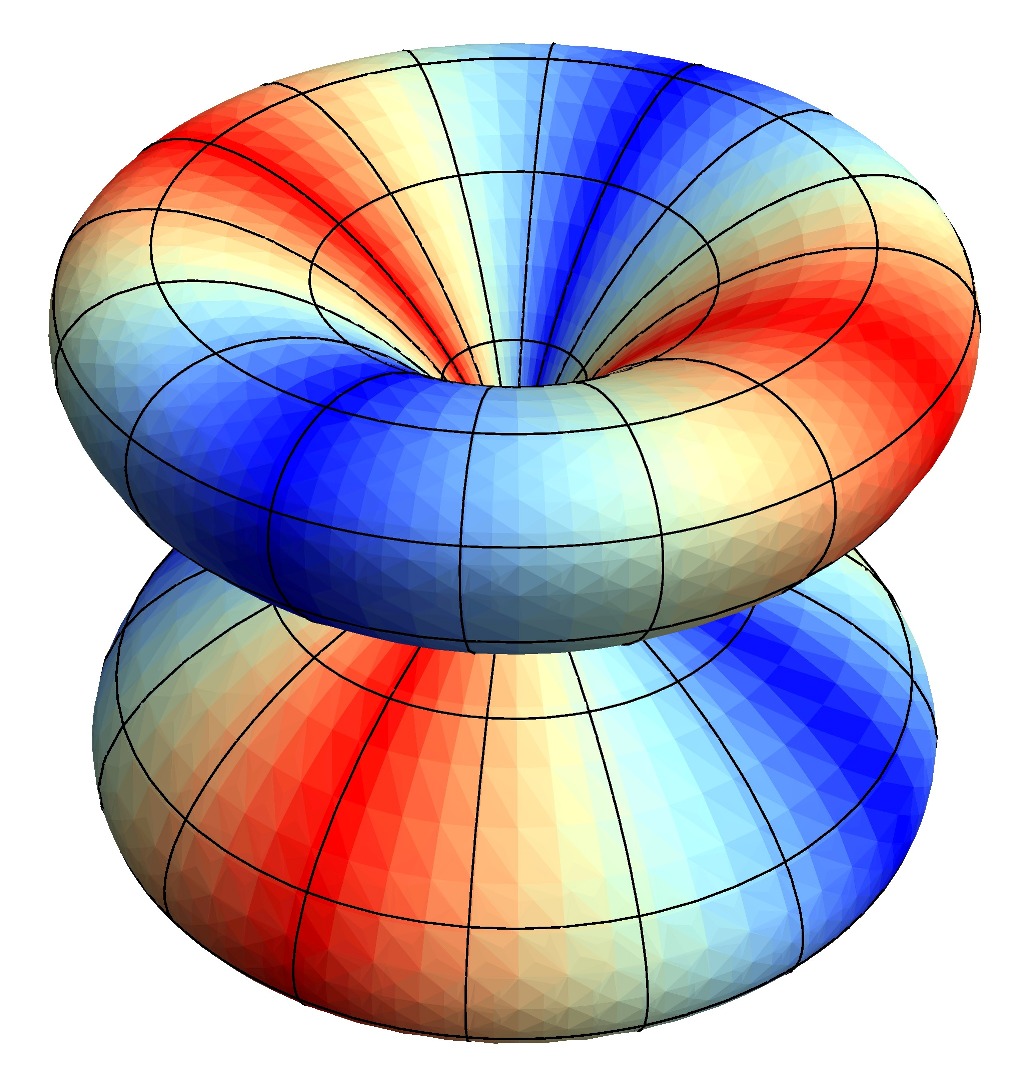}} \\
& & \\
& & \\
& & \\
& & \\
& & \\
& & \\
& $\overline{d}(k)=\Delta(T)(k^{2}_{x}-k^2_y)k_z\widehat{z}$ & $\overline{d}(k)=\Delta(T)(k_x+ik_y)^2k_z\widehat{z}$\\
\hline  

\end{tabular}
\label{OPtable}
\caption{(color online) Graphical depictions of the $E_{1g}$ and $E_{2u}$ models of the order parameter for UPt$_3$.  Columns denote the high and low temperature (A and B, respectively) superconducting phases, and rows denote the two theoretical models.}
\end{table}

Josephson interferometry, used successfully to characterize the cuprates as d-wave and Sr$_2$RuO$_4$ as complex p-wave\cite{wollman95,liu04,kidwingira06}, remains the most definitive phase-sensitive test of the order parameter of unconventional superconductors.  In this technique, a superconducting weak link is created between two superconductors - in our case, a single crystal of UPt$_3$ and a film of the conventional superconductor Pb.  Applying a magnetic field to this junction perpendicular to the current flow creates a phase gradient along the junction that alters the local supercurrent density.  Any intrinsic phase differences arising from the order parameter symmetry will also affect the current density.  The critical current($I_c$) can be given as a function of external flux ($\Phi_{ext}$) and intrinsic phase difference ($\delta$) by the following:

\begin{equation}
\label{cornerjct}
I_c(\Phi_{ext}) = I_0\left|  \frac{\sin\left(\pi \Phi_{ext}/\Phi_0 + \delta/2\right)}{\pi \Phi_{ext}/\Phi_0}  \right|
\end{equation}

\begin{figure}
	\includegraphics[width=0.48\textwidth]{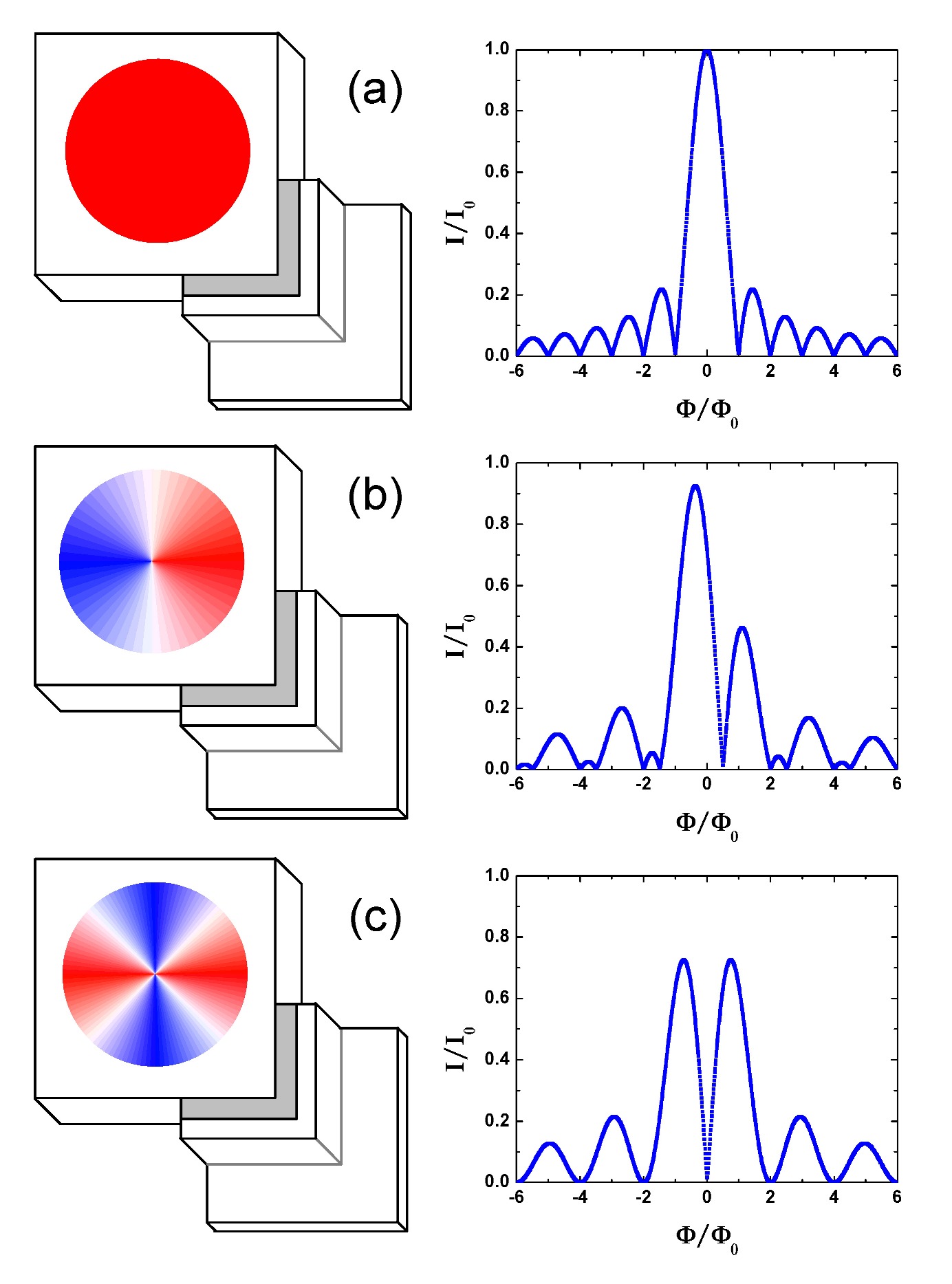}
	\caption{(color online) Planar representations of the order parameter laid on top of a schematic of a corner junction, with the corresponding diffraction pattern placed alongside. (a)An s-wave order parameter produces the classic Fraunhofer pattern. (b)The $E_{1g}$ B-phase produces an asymmetric double peak. (c)The $E_{2u}$ B-phase produces a symmetric double peak.}
	\label{cornerpredict}
\end{figure}

In the case of uniform s-wave superconductors, this results in the conventional Fraunhofer diffraction pattern shape for plots of critical current vs. applied magnetic field.  Even in a superconductor with an anisotropic order parameter, as long as the junction is on a single flat crystal face, the pattern will look Fraunhofer.  This is because tunneling probability falls off exponentially with barrier thickness and so the Josephson current effectively probes a single k-space direction.  If, however, the junction wraps around the corner of a superconductor with a sign change in the order parameter between the two tunneling directions, part of the junction will probe each direction, and the phase shift will cause a distinctive change in the diffraction pattern.  The predicted patterns for each of the order parameter models are given in Figure~\ref{cornerpredict}.

\section{Experiment}
The UPt$_3$ crystals were grown in an electron-beam floating zone furnace, and have a measured residual resistivity ratio (RRR) of greater than 900, in some cases as high as 1100, indicating their exceptional purity.  We polished the surfaces with $0.3 \mu m$ diamond lapping films, and then glued them to a glass substrate with Pyralin\textsuperscript{\textregistered} polyimide coating.  After masking with a dry photoresist, the  were ion milled and $150nm$ of Cu was evaporated as a normal metal barrier, followed by $800nm$ of Pb as the superconducting counter-electrode.  Junction dimensions were typically 50x100$\mu m$.  Junctions on UPt$_3$ are known to be susceptible to magnetic flux trapping\cite{sumiyama05} so the samples were cooled in a Kelvinox\textsuperscript{\textregistered} dilution refrigerator with Cryoperm\textsuperscript{\textregistered} and lead cans to provide the necessary magnetic shielding.  The junction voltages were exceptionally small ($\approx nV$), and so were measured with a superconducting quantum interference device (SQUID) potentiometer circuit, in which an inductively-coupled SQUID detected the current flowing through a known resistor in parallel with the junction.

\begin{figure}
	\includegraphics[width=0.48\textwidth]{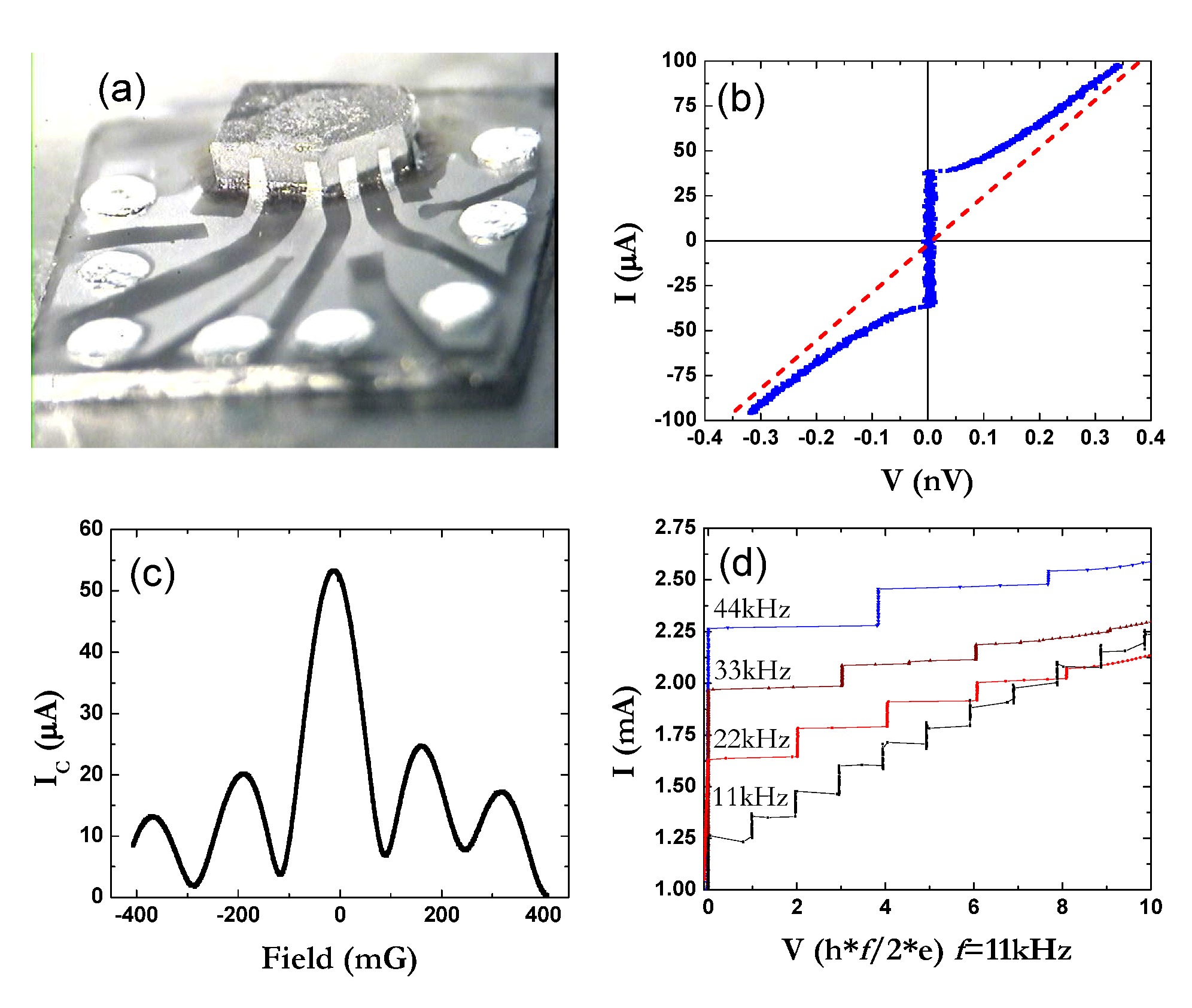}
	\caption{(color online) (a)A photograph of one of our samples - the crystal is the large block in the upper middle with four edge junctions evaporated on its surface.  The lines of Pb film and indium pads are also visible.  (b)A characteristic IV plot for a junction, showing classic RSJ behavior.  (c)A diffraction pattern for one of our edge junctions, exhibiting a nearly Fraunhofer shape.  (d)Shapiro steps from a junction, confirming Josephson behavior.}
	\label{jctcharacter}
\end{figure}

All the measurements reported here were carried out in the low temperature B-phase of UPt$_3$.  The junctions exhibited nearly ideal resistively-shunted junction (RSJ) behavior, as well as showing Shapiro steps when an ac modulation was applied.  Junctions fabricated on a single crystal face displayed diffraction patterns that were nearly Fraunhofer and symmetric around zero field, indicating uniform phase and no trapped vortices.  Examples of these measurements, as well as a sample photo, can be seen in Figure~\ref{jctcharacter}.  It is worth mentioning that even though the B-phase of UPt$_3$ is expected to be chiral and exhibit TRSB, similar to Sr$_2$RuO$_4$, we saw none of the evidence for chiral domains in UPt$_3$ that were seen in Sr$_2$RuO$_4$\cite{kidwingira06}, such as hysteresis or switching noise.
  
We also measured junctions fabricated so that they straddled the corner between the a-and b-axes.  These corner junctions behaved quite differently than the edge junctions.  In all cases the diffraction patterns they produced exhibited features that were asymmetric with respect to field polarity.  The asymmetry was not caused by the self-field effect of the current through the junction, because the pattern was symmetric with respect to the direction of current flow.  Changing temperatures affected the magnitude of the critical current, as expected, but left the shape of the patterns unchanged.  This asymmetry with respect to field polarity is a characteristic sign of a complex order parameter and TRSB, which supports both models' predictions of a chiral state in the B-phase.  

Aside from the ever-present asymmetry, the corner junction patterns varied between different junctions or thermal cycles of the same junction, with three or four qualitatively similar patterns recurring.  The variations in the patterns between different junctions could be due to surface roughness at the corner.  It is unlikely that the two polished surfaces make a perfect corner, and faceting at the region where they meet is probable.  These facets can cause distortions in diffraction patterns, particularly for non-s-wave superconductors, but even in this case an asymmetric variation is a signature of a complex component of the order parameter\cite{hilgenkamp02,neils02}.

\begin{figure}[b]
	\includegraphics[width=0.48\textwidth]{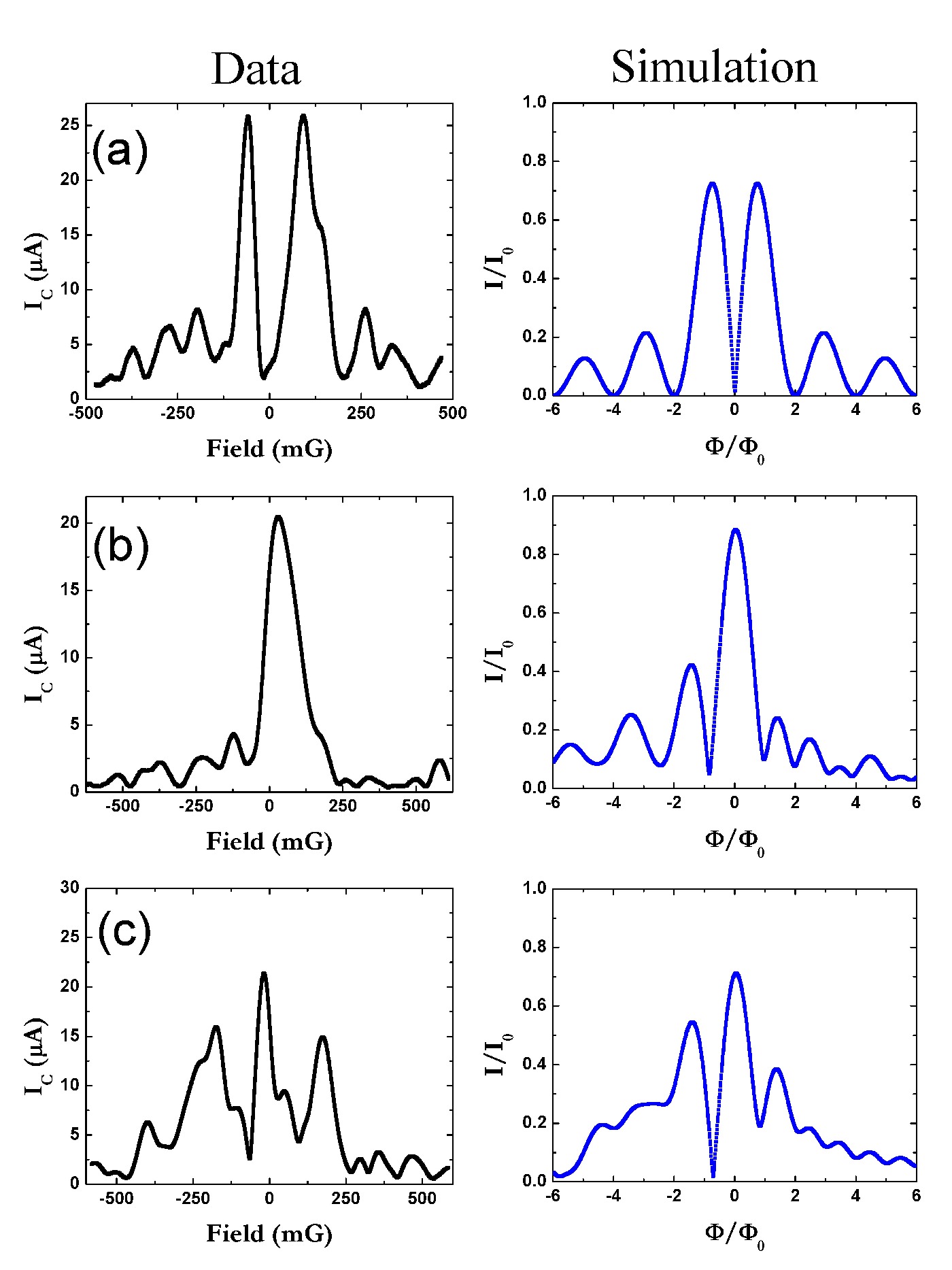}
	\caption{(color online) Comparisons of corner junction diffraction patterns with simulations.  All simulations shown assume the $E_{2u}$ representation.  (a)Corner junction with no vortex trapping.  (b)Junction with a vortex trapped within 5\% of junction width of the corner. (c)Junction with a vortex trapped within 15\% of junction width of the corner.}
	\label{simcompare}
\end{figure}

The changes in the diffraction patterns between thermal cycles require a dynamic mechanism to explain.  The most obvious candidate is vortex trapping near the junctions.  Though our magnetic shielding ($H_{res}\approx10^{-4}G$) combined with very slow cooling cycles seems to be sufficient to prevent vortex entry into the edge junctions, it is possible that the corners of our samples are damaged enough to provide more pinning sites.  With this in mind, we have tried modeling corner junctions combining an intrinsic phase shift with trapped vortices. We tested phase shifts corresponding to the three candidate symmetries: 0 (s-wave), $\pi/2$ ($E_{1g}$), and $\pi$ ($E_{2u}$).  We modeled a vortex as a Gaussian contribution to the flux through the junction with integrated flux = $\Phi_0/2$ and width equal to 3-5\% of the junction width.  We then tried to reproduce the types of patterns we measured most often.  We do not claim to have modeled the junctions exactly, but the best qualitative fit to the observed patterns was given by an intrinsic phase shift of $\pi$ with vortices close to the center of the junction.  Comparisons of a few representative diffraction patterns to simulations are given in Figure~\ref{simcompare}.

In an effort to avoid the complications caused by the material properties of the corners, we fabricated two junctions, one on either side of the corner, incorporating the crystal into a SQUID geometry.  In this case, an intrinsic phase difference in the crystal will show up as a shift in the peak of the critical current modulation, as given by:

\begin{equation}
\label{cornerSQUID}
I_c(\Phi_{ext}) = 2I_0\left|\cos\left(\pi\frac{\Phi_{ext}}{\Phi_0} + \frac{\delta}{2}\right)\right|
\end{equation}

\begin{figure}
\includegraphics[width=0.48\textwidth]{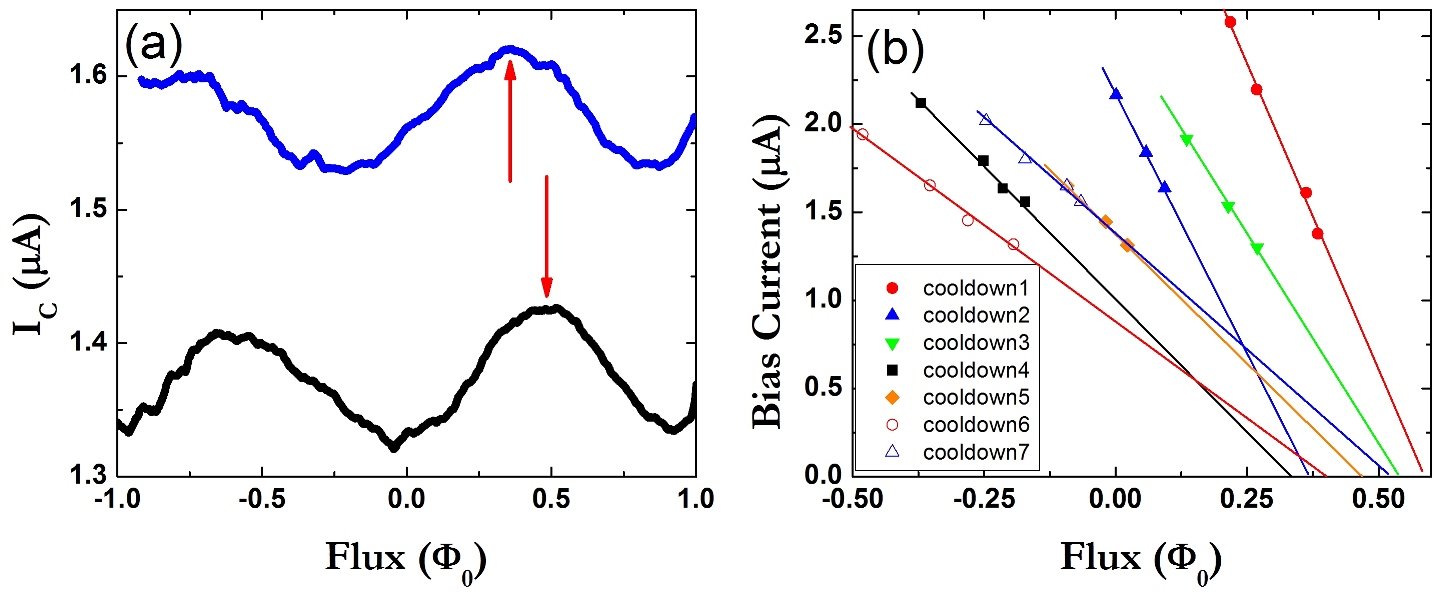}
	\caption{(color online)(a)Two SQUID (one junction on either side of a corner) modulation curves taken at different bias currents - arrows denote the peaks and highlight the shift in position caused by asymmetric current flow.  The arrow locations for these and other curves correspond to data points in plot (b).  (b)Extrapolations to zero bias current for seven thermal cycles of the corner SQUID.  The lines cluster around a phase shift of $0.5\Phi_0$.}
	\label{SQUIDplots}
\end{figure}

Unlike a single junction, the periodic modulation of a SQUID does not provide a central peak to reveal the zero point of magnetic flux, so extra care is required to rule out extrinsic shifts in the pattern.  If the two junctions are not identical, unequal current flow between the branches of the loop will couple field into the SQUID, shifting the pattern.  To account for this, we bias the SQUID at various current levels as shown in Figure~\ref{SQUIDplots}a, noting peak location at each current, and then extrapolate the peak location to zero bias current\cite{wollman93}.  In order to account for residual external field or trapped flux, we performed several thermal cycles, with the results plotted in Figure~\ref{SQUIDplots}b.  The results cluster around $\Phi_0/2$, corresponding to a phase shift of $\pi$, which agrees with our corner junction results and also supports the $E_{2u}$ model.

\section{Conclusions}
In summary, we have fabricated Josephson junctions on high quality single crystals of UPt$_3$ and used them to perform phase sensitive measurements of the superconducting order parameter. There is strong evidence for a complex component of the order parameter from the asymmetry with respect to field polarity in the diffraction patterns, but we found no sign of chiral domains in our edge junctions.  By comparing our corner junction results with simulations involving vortex trapping at the corner and from shifts in SQUID modulation curves, we find evidence for an intrinsic phase shift of $\pi$ for a $90^\circ$ rotation, in agreement with the $E_{2u}$ representation of the order parameter.  We are continuing further measurements with different surface treatments, including as-grown crystal faces, in order to reduce the effect of vortices, as well as studying the crossover between the A and B phases.

The measurements were carried out at the University of Illinois in the Frederick Seitz Materials Research Laboratory and supported by NSF grant DMR07-05214. The UPt$_3$ crystals were grown and annealed at Northwestern University supported by the DOE Basic Energy Sciences grant DE-FG02-05ER46248.  We acknowledge the help of J. Davis, J. Pollanen, H. Choi, T. Lippman, and W. Gannon with crystal growth and characterization.

\bibliography{upt3paper}

\begin{thebibliography}{21}
\expandafter\ifx\csname natexlab\endcsname\relax\def\natexlab#1{#1}\fi
\expandafter\ifx\csname bibnamefont\endcsname\relax
  \def\bibnamefont#1{#1}\fi
\expandafter\ifx\csname bibfnamefont\endcsname\relax
  \def\bibfnamefont#1{#1}\fi
\expandafter\ifx\csname citenamefont\endcsname\relax
  \def\citenamefont#1{#1}\fi
\expandafter\ifx\csname url\endcsname\relax
  \def\url#1{\texttt{#1}}\fi
\expandafter\ifx\csname urlprefix\endcsname\relax\def\urlprefix{URL }\fi
\providecommand{\bibinfo}[2]{#2}
\providecommand{\eprint}[2][]{\url{#2}}

\bibitem[{\citenamefont{Stewart et~al.}(1984)\citenamefont{Stewart, Fisk,
  Willis, and Smith}}]{stewart84}
\bibinfo{author}{\bibfnamefont{G.~R.} \bibnamefont{Stewart}},
  \bibinfo{author}{\bibfnamefont{Z.}~\bibnamefont{Fisk}},
  \bibinfo{author}{\bibfnamefont{J.~O.} \bibnamefont{Willis}},
  \bibnamefont{and} \bibinfo{author}{\bibfnamefont{J.~L.} \bibnamefont{Smith}},
  \bibinfo{journal}{Phys. Rev. Lett.} \textbf{\bibinfo{volume}{52}},
  \bibinfo{pages}{679} (\bibinfo{year}{1984}).

\bibitem[{\citenamefont{Fisher et~al.}(1989)\citenamefont{Fisher, Kim,
  Woodfield, Phillips, Taillefer, Hasselbach, Flouquet, Giorgi, and
  Smith}}]{fisher89}
\bibinfo{author}{\bibfnamefont{R.~A.} \bibnamefont{Fisher}},
  \bibinfo{author}{\bibfnamefont{S.}~\bibnamefont{Kim}},
  \bibinfo{author}{\bibfnamefont{B.~F.} \bibnamefont{Woodfield}},
  \bibinfo{author}{\bibfnamefont{N.~E.} \bibnamefont{Phillips}},
  \bibinfo{author}{\bibfnamefont{L.}~\bibnamefont{Taillefer}},
  \bibinfo{author}{\bibfnamefont{K.}~\bibnamefont{Hasselbach}},
  \bibinfo{author}{\bibfnamefont{J.}~\bibnamefont{Flouquet}},
  \bibinfo{author}{\bibfnamefont{A.~L.} \bibnamefont{Giorgi}},
  \bibnamefont{and} \bibinfo{author}{\bibfnamefont{J.~L.} \bibnamefont{Smith}},
  \bibinfo{journal}{Phys. Rev. Lett.} \textbf{\bibinfo{volume}{62}},
  \bibinfo{pages}{1411} (\bibinfo{year}{1989}).

\bibitem[{\citenamefont{Adenwalla et~al.}(1990)\citenamefont{Adenwalla, Lin,
  Ran, Zhao, Ketterson, Sauls, Taillefer, Hinks, Levy, and
  Sarma}}]{adenwalla90}
\bibinfo{author}{\bibfnamefont{S.}~\bibnamefont{Adenwalla}},
  \bibinfo{author}{\bibfnamefont{S.~W.} \bibnamefont{Lin}},
  \bibinfo{author}{\bibfnamefont{Q.~Z.} \bibnamefont{Ran}},
  \bibinfo{author}{\bibfnamefont{Z.}~\bibnamefont{Zhao}},
  \bibinfo{author}{\bibfnamefont{J.~B.} \bibnamefont{Ketterson}},
  \bibinfo{author}{\bibfnamefont{J.~A.} \bibnamefont{Sauls}},
  \bibinfo{author}{\bibfnamefont{L.}~\bibnamefont{Taillefer}},
  \bibinfo{author}{\bibfnamefont{D.~G.} \bibnamefont{Hinks}},
  \bibinfo{author}{\bibfnamefont{M.}~\bibnamefont{Levy}}, \bibnamefont{and}
  \bibinfo{author}{\bibfnamefont{B.~K.} \bibnamefont{Sarma}},
  \bibinfo{journal}{Phys. Rev. Lett.} \textbf{\bibinfo{volume}{65}},
  \bibinfo{pages}{2298} (\bibinfo{year}{1990}).

\bibitem[{\citenamefont{Behnia et~al.}(1991)\citenamefont{Behnia, Taillefer,
  Flouquet, Jaccard, Maki, and Fisk}}]{behnia91}
\bibinfo{author}{\bibfnamefont{K.}~\bibnamefont{Behnia}},
  \bibinfo{author}{\bibfnamefont{L.}~\bibnamefont{Taillefer}},
  \bibinfo{author}{\bibfnamefont{J.}~\bibnamefont{Flouquet}},
  \bibinfo{author}{\bibfnamefont{D.}~\bibnamefont{Jaccard}},
  \bibinfo{author}{\bibfnamefont{K.}~\bibnamefont{Maki}}, \bibnamefont{and}
  \bibinfo{author}{\bibfnamefont{Z.}~\bibnamefont{Fisk}}, \bibinfo{journal}{J.
  Low Temp. Phys.} \textbf{\bibinfo{volume}{84}}, \bibinfo{pages}{261}
  (\bibinfo{year}{1991}).

\bibitem[{\citenamefont{Suderow et~al.}(1997)\citenamefont{Suderow, Brison,
  Huxley, and Flouquet}}]{suderow97}
\bibinfo{author}{\bibfnamefont{H.}~\bibnamefont{Suderow}},
  \bibinfo{author}{\bibfnamefont{J.~P.} \bibnamefont{Brison}},
  \bibinfo{author}{\bibfnamefont{A.}~\bibnamefont{Huxley}}, \bibnamefont{and}
  \bibinfo{author}{\bibfnamefont{J.}~\bibnamefont{Flouquet}},
  \bibinfo{journal}{J. Low Temp. Phys.} \textbf{\bibinfo{volume}{108}},
  \bibinfo{pages}{11} (\bibinfo{year}{1997}).

\bibitem[{\citenamefont{Luke et~al.}(1993)\citenamefont{Luke, Keren, Le, Wu,
  Uemura, Bonn, Taillefer, and Garrett}}]{luke93}
\bibinfo{author}{\bibfnamefont{G.~M.} \bibnamefont{Luke}},
  \bibinfo{author}{\bibfnamefont{A.}~\bibnamefont{Keren}},
  \bibinfo{author}{\bibfnamefont{L.~P.} \bibnamefont{Le}},
  \bibinfo{author}{\bibfnamefont{W.~D.} \bibnamefont{Wu}},
  \bibinfo{author}{\bibfnamefont{Y.~J.} \bibnamefont{Uemura}},
  \bibinfo{author}{\bibfnamefont{D.~A.} \bibnamefont{Bonn}},
  \bibinfo{author}{\bibfnamefont{L.}~\bibnamefont{Taillefer}},
  \bibnamefont{and} \bibinfo{author}{\bibfnamefont{J.~D.}
  \bibnamefont{Garrett}}, \bibinfo{journal}{Phys. Rev. Lett.}
  \textbf{\bibinfo{volume}{71}}, \bibinfo{pages}{1466} (\bibinfo{year}{1993}).

\bibitem[{\citenamefont{de~Reotier et~al.}(1995)\citenamefont{de~Reotier,
  Huxley, Yaouanc, Flouquet, Bonville, Imbert, Pari, Gubbens, and
  Mulders}}]{reotier95}
\bibinfo{author}{\bibfnamefont{P.~D.} \bibnamefont{de~Reotier}},
  \bibinfo{author}{\bibfnamefont{A.}~\bibnamefont{Huxley}},
  \bibinfo{author}{\bibfnamefont{A.}~\bibnamefont{Yaouanc}},
  \bibinfo{author}{\bibfnamefont{J.}~\bibnamefont{Flouquet}},
  \bibinfo{author}{\bibfnamefont{P.}~\bibnamefont{Bonville}},
  \bibinfo{author}{\bibfnamefont{P.}~\bibnamefont{Imbert}},
  \bibinfo{author}{\bibfnamefont{P.}~\bibnamefont{Pari}},
  \bibinfo{author}{\bibfnamefont{P.}~\bibnamefont{Gubbens}}, \bibnamefont{and}
  \bibinfo{author}{\bibfnamefont{A.}~\bibnamefont{Mulders}},
  \bibinfo{journal}{Phys. Lett. A} \textbf{\bibinfo{volume}{205}},
  \bibinfo{pages}{239} (\bibinfo{year}{1995}).

\bibitem[{\citenamefont{Tou et~al.}(1996)\citenamefont{Tou, Kitaoka, Asayama,
  Kimura, Onuki, Yamamoto, and Maezawa}}]{tou96}
\bibinfo{author}{\bibfnamefont{H.}~\bibnamefont{Tou}},
  \bibinfo{author}{\bibfnamefont{Y.}~\bibnamefont{Kitaoka}},
  \bibinfo{author}{\bibfnamefont{K.}~\bibnamefont{Asayama}},
  \bibinfo{author}{\bibfnamefont{N.}~\bibnamefont{Kimura}},
  \bibinfo{author}{\bibfnamefont{Y.}~\bibnamefont{Onuki}},
  \bibinfo{author}{\bibfnamefont{E.}~\bibnamefont{Yamamoto}}, \bibnamefont{and}
  \bibinfo{author}{\bibfnamefont{K.}~\bibnamefont{Maezawa}},
  \bibinfo{journal}{Phys. Rev. Lett.} \textbf{\bibinfo{volume}{77}},
  \bibinfo{pages}{1374} (\bibinfo{year}{1996}).

\bibitem[{\citenamefont{Park and Joynt}(1996)}]{park96}
\bibinfo{author}{\bibfnamefont{K.~A.} \bibnamefont{Park}} \bibnamefont{and}
  \bibinfo{author}{\bibfnamefont{R.}~\bibnamefont{Joynt}},
  \bibinfo{journal}{Phys. Rev. B} \textbf{\bibinfo{volume}{53}},
  \bibinfo{pages}{12346} (\bibinfo{year}{1996}).

\bibitem[{\citenamefont{Sauls}(1994)}]{sauls94}
\bibinfo{author}{\bibfnamefont{J.~A.} \bibnamefont{Sauls}},
  \bibinfo{journal}{Adv. in Phys.} \textbf{\bibinfo{volume}{43}},
  \bibinfo{pages}{113} (\bibinfo{year}{1994}).

\bibitem[{\citenamefont{Norman and Hirschfeld}(1996)}]{norman96}
\bibinfo{author}{\bibfnamefont{M.~R.} \bibnamefont{Norman}} \bibnamefont{and}
  \bibinfo{author}{\bibfnamefont{P.~J.} \bibnamefont{Hirschfeld}},
  \bibinfo{journal}{Phys. Rev. B} \textbf{\bibinfo{volume}{53}},
  \bibinfo{pages}{5706} (\bibinfo{year}{1996}).

\bibitem[{\citenamefont{Graf et~al.}(2000)\citenamefont{Graf, Yip, and
  Sauls}}]{graf00}
\bibinfo{author}{\bibfnamefont{M.~J.} \bibnamefont{Graf}},
  \bibinfo{author}{\bibfnamefont{S.}~\bibnamefont{Yip}}, \bibnamefont{and}
  \bibinfo{author}{\bibfnamefont{J.~A.} \bibnamefont{Sauls}},
  \bibinfo{journal}{Phys. Rev. B} \textbf{\bibinfo{volume}{62}},
  \bibinfo{pages}{14393} (\bibinfo{year}{2000}).

\bibitem[{\citenamefont{Wu and Joynt}(2002)}]{wu02}
\bibinfo{author}{\bibfnamefont{W.~C.} \bibnamefont{Wu}} \bibnamefont{and}
  \bibinfo{author}{\bibfnamefont{R.}~\bibnamefont{Joynt}},
  \bibinfo{journal}{Phys. Rev. B} \textbf{\bibinfo{volume}{65}},
  \bibinfo{pages}{104502} (\bibinfo{year}{2002}).

\bibitem[{\citenamefont{Joynt and Taillefer}(2002)}]{joynt02}
\bibinfo{author}{\bibfnamefont{R.}~\bibnamefont{Joynt}} \bibnamefont{and}
  \bibinfo{author}{\bibfnamefont{L.}~\bibnamefont{Taillefer}},
  \bibinfo{journal}{Rev. Mod. Phys.} \textbf{\bibinfo{volume}{74}},
  \bibinfo{pages}{235} (\bibinfo{year}{2002}).

\bibitem[{\citenamefont{Wollman et~al.}(1995)\citenamefont{Wollman, {Van
  Harlingen}, Giapintzakis, and Ginsberg}}]{wollman95}
\bibinfo{author}{\bibfnamefont{D.~A.} \bibnamefont{Wollman}},
  \bibinfo{author}{\bibfnamefont{D.~J.} \bibnamefont{{Van Harlingen}}},
  \bibinfo{author}{\bibfnamefont{J.}~\bibnamefont{Giapintzakis}},
  \bibnamefont{and} \bibinfo{author}{\bibfnamefont{D.~M.}
  \bibnamefont{Ginsberg}}, \bibinfo{journal}{Phys. Rev. Lett.}
  \textbf{\bibinfo{volume}{74}}, \bibinfo{pages}{797} (\bibinfo{year}{1995}).

\bibitem[{\citenamefont{Nelson et~al.}(2004)\citenamefont{Nelson, Mao, Maeno,
  and Liu}}]{liu04}
\bibinfo{author}{\bibfnamefont{K.~D.} \bibnamefont{Nelson}},
  \bibinfo{author}{\bibfnamefont{Z.~Q.} \bibnamefont{Mao}},
  \bibinfo{author}{\bibfnamefont{Y.}~\bibnamefont{Maeno}}, \bibnamefont{and}
  \bibinfo{author}{\bibfnamefont{Y.}~\bibnamefont{Liu}},
  \bibinfo{journal}{Science} \textbf{\bibinfo{volume}{306}},
  \bibinfo{pages}{1151} (\bibinfo{year}{2004}).

\bibitem[{\citenamefont{Kidwingira et~al.}(2006)\citenamefont{Kidwingira,
  Strand, {Van Harlingen}, and Maeno}}]{kidwingira06}
\bibinfo{author}{\bibfnamefont{F.}~\bibnamefont{Kidwingira}},
  \bibinfo{author}{\bibfnamefont{J.~D.} \bibnamefont{Strand}},
  \bibinfo{author}{\bibfnamefont{D.~J.} \bibnamefont{{Van Harlingen}}},
  \bibnamefont{and} \bibinfo{author}{\bibfnamefont{Y.}~\bibnamefont{Maeno}},
  \bibinfo{journal}{Science} \textbf{\bibinfo{volume}{314}},
  \bibinfo{pages}{1267} (\bibinfo{year}{2006}).

\bibitem[{\citenamefont{Sumiyama et~al.}(2005)\citenamefont{Sumiyama, Hata,
  Oda, Kimura, Yamamoto, Haga, and Onuki}}]{sumiyama05}
\bibinfo{author}{\bibfnamefont{A.}~\bibnamefont{Sumiyama}},
  \bibinfo{author}{\bibfnamefont{R.}~\bibnamefont{Hata}},
  \bibinfo{author}{\bibfnamefont{Y.}~\bibnamefont{Oda}},
  \bibinfo{author}{\bibfnamefont{N.}~\bibnamefont{Kimura}},
  \bibinfo{author}{\bibfnamefont{E.}~\bibnamefont{Yamamoto}},
  \bibinfo{author}{\bibfnamefont{Y.}~\bibnamefont{Haga}}, \bibnamefont{and}
  \bibinfo{author}{\bibfnamefont{Y.}~\bibnamefont{Onuki}},
  \bibinfo{journal}{Phys. Rev. B} \textbf{\bibinfo{volume}{72}},
  \bibinfo{pages}{174507} (\bibinfo{year}{2005}).

\bibitem[{\citenamefont{Hilgenkamp and Mannhart}(2002)}]{hilgenkamp02}
\bibinfo{author}{\bibfnamefont{H.}~\bibnamefont{Hilgenkamp}} \bibnamefont{and}
  \bibinfo{author}{\bibfnamefont{J.}~\bibnamefont{Mannhart}},
  \bibinfo{journal}{Rev. Mod. Phys.} \textbf{\bibinfo{volume}{74}},
  \bibinfo{pages}{485} (\bibinfo{year}{2002}).

\bibitem[{\citenamefont{Neils and {Van Harlingen}}(2002)}]{neils02}
\bibinfo{author}{\bibfnamefont{W.~K.} \bibnamefont{Neils}} \bibnamefont{and}
  \bibinfo{author}{\bibfnamefont{D.~J.} \bibnamefont{{Van Harlingen}}},
  \bibinfo{journal}{Phys. Rev. Lett.} \textbf{\bibinfo{volume}{88}},
  \bibinfo{pages}{47001} (\bibinfo{year}{2002}).

\bibitem[{\citenamefont{Wollman et~al.}(1993)\citenamefont{Wollman, {Van
  Harlingen}, Lee, Ginsberg, and Leggett}}]{wollman93}
\bibinfo{author}{\bibfnamefont{D.~A.} \bibnamefont{Wollman}},
  \bibinfo{author}{\bibfnamefont{D.~J.} \bibnamefont{{Van Harlingen}}},
  \bibinfo{author}{\bibfnamefont{W.~C.} \bibnamefont{Lee}},
  \bibinfo{author}{\bibfnamefont{D.~M.} \bibnamefont{Ginsberg}},
  \bibnamefont{and} \bibinfo{author}{\bibfnamefont{A.~J.}
  \bibnamefont{Leggett}}, \bibinfo{journal}{Phys. Rev. Lett.}
  \textbf{\bibinfo{volume}{71}}, \bibinfo{pages}{2134} (\bibinfo{year}{1993}).

\end{thebibliography}
\end{document}